\documentclass[5p,preprint]{elsarticle}
\usepackage{microtype}

\journal{Physics Letters B}

\newcommand{\im}{\textsc{im}}

\begin{document}

\begin{frontmatter}
\title{Entropy development in ideal relativistic fluid dynamics with the Bag Model equation of state}

\author[uib,ubb]{Sz.~Horv\'at}
\ead{szhorvat@ift.uib.no}
\author[ub]{V.~K.~Magas}
\author[fias,uv]{D.~D.~Strottman}
\author[uib,fias,kfki]{L.~P.~Csernai}
%\ead{csernai@ift.uib.no}

\address[uib]{Institute of Physics and Technology, University of Bergen, All\'egaten 55, N-5007 Bergen, Norway}

\address[ubb]{Babe\c{s}-Bolyai University, Dept.\ of Physics, Kog\u{a}lniceanu street 1, RO-400084 Cluj-Napoca, Romania}

\address[ub]{Departament d'Estructura i Constituents de la Mat\`eria, Universitat de Barcelona, Diagonal 647, 08028 Barcelona, Spain}

\address[fias]{Frankfurt Institute for Advanced Studies, J.W. Goethe University, Ruth-Moufgang-Str.~1, 60438 Frankfurt am Main, Germany}

\address[uv]{Dep. de Fisica Teorica and IFIC, Centro Mixto Universidad de Valencia \\
CSIC, Institutos de Investigaci\'on de Paterna, Aptd. 22085, 46071 Valencia, Spain}

\address[kfki]{MTA-KFKI, Research Inst.~of Particle and Nuclear Physics, 1525 Budapest, Hungary}

\begin{abstract}
We consider an idealized situation where the Quark-Gluon Plasma
(QGP) is described by a perfect, 3+1 dimensional fluid dynamic model starting from an initial state and expanding until a final state where freeze-out and/or hadronization takes place. We study the entropy production with attention to effects of (i) numerical viscosity, (ii) late stages of flow where the Bag Constant and the partonic pressure are becoming similar, (iii) and the consequences of final freeze-out and constituent quark matter formation.
\end{abstract}

\end{frontmatter}

\section{Introduction}

Experimental observations of increased collective flow in ultra-relativistic
heavy ion collisions at CERN SPS and BNL RHIC, as well as theoretical considerations \cite{kovtun_viscosity_2005} suggest that the Quark-Gluon Plasma (QGP) is nearly a perfect fluid with very small viscosity.  Furthermore, it was pointed out \cite{csernai_kapusta_mclerran_2006} that shear viscosity is minimal in fluids just at the critical point of the phase transition, so measuring the minimal viscosity may lead to finding the critical point of the phase transition between QGP and hadronic matter.  Studying the entropy development of QGP systems provides an important aid in this.

A fluid dynamical (FD) description of the nuclear matter will be considered here with the example of an Au + Au central collision at $65+65$ A$\cdot$GeV.  Using the MIT Bag Model Equation of State (EoS), we consider the physical vacuum to have zero energy, except in the volume domain where QGP exists.  Initially this volume is $V_{\mathrm{init}}$.  Then the matter expands in the flow and reaches a larger freeze-out (FO) volume, $V_{FO}$.  A FD description does not constrain the location of the FO.  Usually external conditions are given to determine it, such as a prescribed FO temperature or zero pressure, etc.  The fluid dynamical model we use \cite{csernai_collective_flow_2010,csernai_flow_analysis_2009,bravina_collective_dynamics_1998} can therefore run well beyond the FO, so that the location of physical FO can be selected afterwards as a space-time hyper-surface or layer, where FO occurs.  In order to select the most realistic FO conditions, the temperature and entropy density of the fluid may provide important guidance, although the FD model does not use these quantities directly, only the pressure and the energy density (and eventually the net baryon density) in the EoS.

\begin{figure}[thb]
  \begin{center}
    \includegraphics{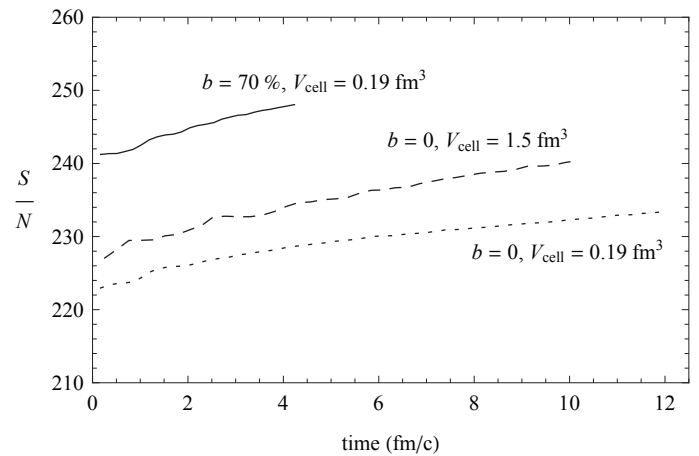}
  \end{center}
  \caption{The mean specific entropy, $S/N$, is shown for three different FD computations ($N$ is the number of participants).  Although the simulations were done for adiabatic expansion of an ideal fluid, the entropy increases due to the numerical viscosity of the method.  The difference in initial specific entropy between the two cases describing collisions with impact parameter $b=0$ is due to coarse graining. $V_{\mathrm{cell}}$ denotes the cell size of the computational grid.}
  \label{fig:entropy1}
\end{figure}

The 3+1 dimensional, FD model \cite{csernai_collective_flow_2010,csernai_flow_analysis_2009,bravina_collective_dynamics_1998} uses the Particle in Cell (PIC) method adapted to ultra-relativistic heavy ion collisions. In his method, so called marker particles, which are smaller Lagrangian fluid cells corresponding to fixed baryon charge, move in a grid of larger Eulerian cells.  The computational fluid dynamical (CFD) calculation, describing the participant zone of the reaction containing a net baryon charge N, starts from an analytic initial state model \cite{magas_initial_state_2001,magas_initial_state_2002} based on longitudinally expanding strings of the color-magnetic field allowing for different string tensions, string cross-sections and for finite impact parameters. The entropy of the initial state depends on the resolution of the spatial grid, as the coarse graining with the relatively small initial spatial size leads to an observable difference (see figure~\ref{fig:entropy1}). The initial entropy is substantial, it is in agreement with other estimates, e.g. \cite{FriesNPA09}. The subsequent expansion shows a weak entropy increase, 5-6 \%, due to the numerical viscosity, although the model considers a perfect fluid. The entropy increase due to numerical viscosity is smaller when the cell size is smaller. At late stages the entropy increase is weaker due to the smaller gradients between neighbouring cells in the expanding volume. The entropy increase due to numerical viscosity exceeds the one obtained in 1+1 and 2+1 dimensional models with smaller cell sizes \cite{SHB08}, and it is somewhat below the entropy increase of 6-24 \%, obtained in those models due to physical viscosity.

The MIT Bag Model EoS yields a negative pressure at low energy densities (which occur at late stages of the expansion) indicating a tendency for clusterization.  As in most CFD approaches, the pressure is cut at zero, i.e.\ no negative pressure is allowed in the numerical calculation.  This is necessary for the stable calculation of the late expansion stages.  This choice also ensures that the regions past the FO surface, most of which will have zero pressure in the FD calculation, have a minimal influence on the regions behind it, as these cells exercise no work, $pdV$, on their neighbours.

The MIT Bag Model assumes two components: an ideal parton gas and a perturbative vacuum.  The perturbative vacuum has a vacuum energy, $BV$, where $B$ is the bag constant (in our calculations $B = 0.396$ GeV/fm$^3$), and $V$ is the volume of the QGP.  Thus, this energy is increasing as the plasma expands.

This energy increase happens at the expense of the work done by the partonic pressure, $p_p dV$, which partly accelerates the neighbouring cells (causing the observed flow), and is partly invested into the increased energy of the Bag. It is assumed that the parton gas is an ideal Stefan--Boltzmann gas, where $p_p = e_p/3$ so the Bag model EoS  with $e = e_p + B$ and $p = p_p - B$ becomes: $p = e/3 - 4B/3$.  

\section{Final stages of the fluid dynamical expansion}

Since the late stages of the expansion are important for the subsequent description of hadronization and freeze-out, we scrutinize the temperature and entropy development at these stages in FD.  For transparency, we use the MIT Bag Model EoS, and compare it qualitatively to more realistic EoS features from Lattice QCD.

When the total pressure drops to zero then the total $p dV$ is also zero.  This usually happens in surface cells, which continue to participate in the expansion due to their inertia.  At the same time, if the energy density of the Bag remains constant, $B$, the total energy of the Bag increases further.  It must be noted that all these calculations were done with the assumption that the Bag field is uniform and homogeneous, and thus its entropy vanishes.  Obviously, near the hadronization and FO the applicability of the MIT Bag Model EoS is overstreched.  The Bag field should  gradually disappear and hadrons will be formed.  The assumption of an equilibrium phase mixture \cite{csernai_kapusta_dynamics_1992,csernai_kapusta_nucleation_1992} leads to a slow hadronization, which is not supported by the spacetime extent indicated by two particle correlations.  An alternative supercooling followed by fast and simultaneous hadronization and FO is advocated for %??
in refs.~\cite{csorgo_csernai_freezeout_1994,csernai_mishustin_fast_1995,zetenyi2010}.  
We apply this approach so that the FD stage of the model concludes in the final simultaneous hadronization and freeze-out. %%??

We shall study the details of this process.  
To have a better insight we shall consider the Bag (i.e.\ background field energy) and the parton gas as two separate components of a coupled system.
%It can be shown that if the pressure is simply taken to be 0 where the equation $p = e/3 - 4B/3$ would give a negative value, then it is not possible to describe the QGP as a single thermodynamic system possessing a consistent equation of state.  This is due to the jump in the derivative of the pressure with respect to energy density at the point $e=4B$.  Therefore we shall consider the Bag (i.e.\ background field energy) and the parton gas as two non-thermally coupled components of a composite system.

To derive the thermodynamic properties of the QGP, such as temperature or entropy density, it is necessary to decompose the total energy density $e$ provided by the FD calculation into the energy density of the parton gas ($e_p$) and the energy density of the Bag ($e_B$): $e = e_p + e_B$.  Then thermodynamic quantities can be obtained from the EoS of the parton gas.  Doing this decomposition is trivial when the pressure if strictly positive and $e_B = B$ is a constant.  However, when $p=0$, several choices are possible: Since the system is described as consisting of two parts, and cannot be treated as a single thermodynamic system in equilibrium, the decomposition can only be done unambiguously if (1) a particular change in state is assumed and (2) explicit assumptions are made about the non-mechanical energy exchange between the two components (the Bag field and the parton gas).

Since the FD calculation assumes an ideal fluid, the change in state is ideally adiabatic for the whole (coupled) system, i.e.\ once the pressure $p$ has reached zero, the work done on one Lagrangian cell by its neighbours vanishes, $pdV=0$, and the energy $E = E_p + E_B$ of a Lagrangian fluid cell will stay constant during the expansion.  For the energy exchange between the two sub-systems within one cell (Bag field and parton gas) several choices are possible when the pressure becomes zero.  For simplicity, let us study a baryon-free Stefan-Boltzmann gas, and consider the following possibilities:

\begin{figure}
  \begin{center}
    \includegraphics{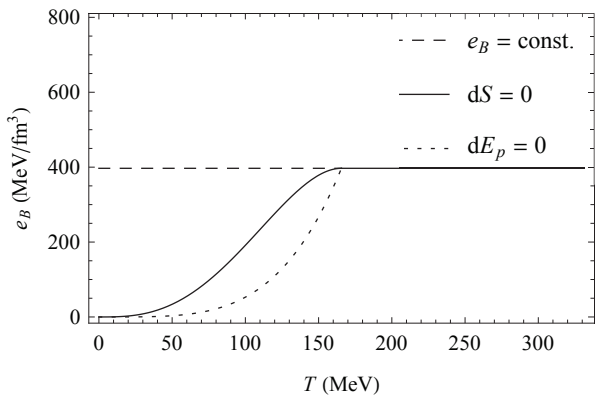}
  \end{center}
  \caption{The energy density of the Bag field, $e_B$, as a function of the temperature of the partonic gas, $T$, for adiabatic expansion (full line, $dS=0$), constant energy of the Bag field (dashed line), and without change of the energy of the parton gas (dotted line).  Case $dE_p = 0$ assumes that the total energy of the ideal gas component remains constant, thus the total energy of the Bag field remains constant as well during the final expansion of the system, while its energy density decreases.  Realistically, the Bag field must disappear in the transition, so the real trajectory must lay below the dotted line.  This disappearance of the Bag field may start before the total pressure drops to zero.}
  \label{fig:eb}
\end{figure}

The \emph{first} possibility is assuming that the energy density, $e_B$, of the Bag field stays constant during the expansion, $e_B = B = \textrm{const}$.  The energy of the Bag field, $BV$, increases strongly during the expansion.  This is only possible if energy is transferred non-mechanically (i.e.\ in another way than through the work done by the partonic pressure $p_p$) from the parton gas component to the Bag, and thus the entropy of the parton gas decreases. 

A \emph{second} possibility is assuming that the only energy exchange between the two sub-systems is due to mechanical interaction (the work $p_p dV$, and thus the entropy of the parton gas is constant, $dS_p = 0$.   In an adiabatic expansion  of the Stefan-Boltzmann gas starting from $V_0$, the energy density decreases $e_p/e_{p0} = (V/V_0)^{-4/3}$, where $e_{p0}=3B$ when the total pressure reaches zero. As the total energy is conserved $(e_p + e_B) V = (e_{p0}+B) V_0$, the volume ratio can be expressed via the energy density ratio, so $(e_p(T)/3B)=\bigl[\bigl(e_p(T)+e_B(T)\bigr)/(4B)\bigr]^{4/3}$. From here $e_B(T)$ can be expressed as shown in figure~\ref{fig:eb} 

The \emph{third} choice considered here is that the energy of the parton gas in a Lagrangian cell stays constant after the pressure has reached zero: $E_p = \textrm{const.}$, or $dE_p = 0$.  This is possible if in addition to the mechanical work that the parton gas does on the Bag field, it also draws energy from it non-mechanically, resulting in a zero net energy exchange.  Thus the entropy of the parton gas is increasing, i.e.\ this is a dissipative expansion.

Finally, if the net energy transfer is directed from the Bag field to the parton gas, then the entropy increase is even larger.  This possibility is the  physically most realistic one.

For these cases, the energy density, $e_B$, of the Bag field is plotted as a function of the temperature of the parton gas for all three cases in figure~\ref{fig:eb}.

\begin{figure}
  \begin{center}
    \includegraphics{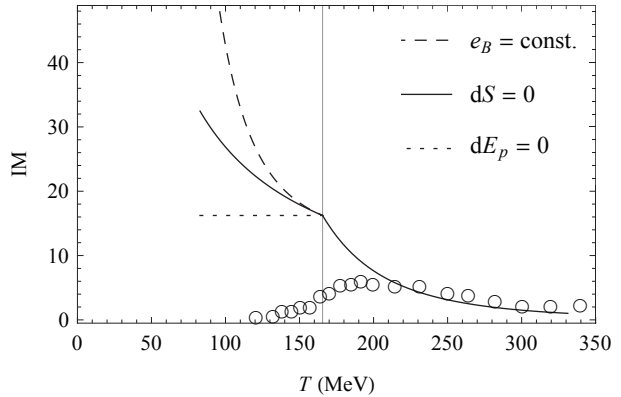}
  \end{center}
  \caption{The interaction measure, \im,  as a function of temperature for the same three cases as in fig.~\ref{fig:eb}.  These curves are compared with the interaction measure obtained from a lattice QCD calculation, represented by empty circles.  There is relatively good agreement between the prediction of the Bag Model and that of the lattice QCD calculations above a temperature of 200 MeV. \cite{milc_2005}  In a more realistic situation when the Bag field diappears in the final expansion, the trajectory would lay below the dotted line.}
  \label{fig:im}
\end{figure}

For the purpose of comparison with Lattice QCD simulations, the interaction measure, $\im = (\hbar c)^3 (e - 3p)/T^4$, can also be computed for these three cases.  The results are plotted in figure~\ref{fig:im}, and compared with the prediction of a lattice QCD calculation. \cite{milc_2005}

In the numerical FD model after about 3-4 fm/c time some fluid cells reach the zero pressure, and as time progresses the number of such cells is increasing.  The zero pressure cells appear early in the calculation because the initial state assumes a smooth surface with low density for cells at the surface of the fluid.  Although the FD model development is only governed by the pressure, the entropy and the temperature development after this time depends on the assumptions of the energy transfer between the homogeneous Bag field and the parton gas.  According to the first choice the entropy of the system may remain constant and then energy should be transferred from the Bag field to the parton gas.

\begin{figure}[h]
  \begin{center}
    \includegraphics{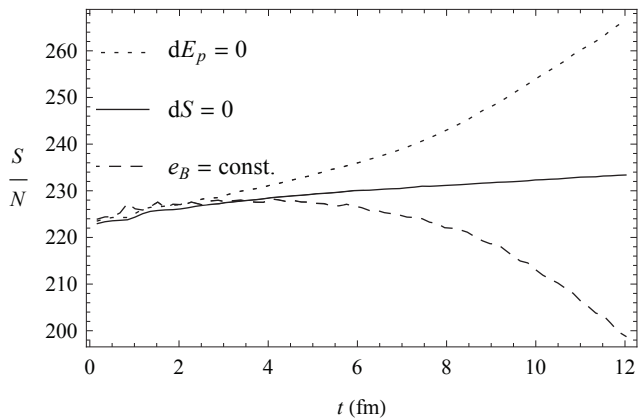}
  \end{center}
  \caption{Results for an Au + Au collision at $65+65$ A$\cdot$GeV energy at impact parameter $b=0$, from a CFD calculation with the Particle in Cell (PiC) method with cell size $dx = dy = dz = 0.575 \;\mathrm{fm}$.  The mean specific entropy of the Au + Au system, $S/N$, as a function of time in the numerical fluid dynamics simulation of a heavy ion collision.  Solid line: adiabatic expansion of the ideal gas component, dashed line: $e_B = B = \textrm{const.}$, dotted line: $E_p = \textrm{const.}$ The slight entropy increase in the ``adiabatic'' case is due to numerical viscosity.}
  \label{fig:entropy}
\end{figure}

A FD simulation was run for the collision of two Au nuclei at $65+65$ A$\cdot$GeV energy and impact parameter $b=0$.  The entropy evolution of the system is plotted for all three cases (see figure~\ref{fig:entropy}).  The
%slight 
entropy increase in the first case (solid line) is due to the numerical viscosity of the CFD model.

The $e_B = \textrm{const.}$ choice leads to an entropy decrease.  The second law of thermodynamics rules out the possibility of the entropy decreasing, therefore the energy density of the Bag field cannot be constant.  More generally, energy cannot be transferred non-mechanically from the parton gas to the unifrom and homogeneous Bag field, and so all such possibilities with total entropy decrease are excluded.  However, energy transfer in the other direction is allowed, as this corresponds to an increase in entropy.

The final hadronization and freeze-out may well proceed through a Quarkyonic matter phase \cite{mclerran_pisarski_2007,hidaka_phase_diagram_2008}, which suggests that first the deconfinement ceases (i.e.\ the Bag field fragments in our picture), and then the chiral symmetry will be broken, so that the current quarks gain mass and become constituent quarks.  It is suggested \cite{csernai_collective_flow_2010} that the Constituent Quark Scaling \cite{adams_2004,adare_2007} may provide a possibility to extract information from the Quarkyonic Matter state.

The present calculation indicates that in the dynamics of this development and the disappearance of the Bag field the entropy plays an important role, especially the entropy arising from the structure of the breaking up Bag field.

Due to the interaction of gluons, we can anticipate that in a given domain of space, the energy density of the Bag may only be $e_B = B$ (if the field is present), or $e_B = 0$ (if it is not).  Thus an average decrease in the energy density of the Bag field, $e_B$, corresponds to the field breaking up into smaller pieces.  When this happens, the field will not be uniform and homogeneous any more, therefore its entropy will need to be taken into account.  The present study provides valuable guidance, indicating the possibilities of the detailed description of the break-up of the background field and the subsequent breaking of the chiral symmetry.

\end{document}